\documentclass[manuscript]{aastex}
\usepackage{tikz}
\usepackage{amsmath}

\usepackage{graphicx}

\usepackage{pdflscape}

\usepackage{wrapfig}

\slugcomment{\textbf{Submitted 2023 June 4; 
 Revised June 25} }

\shorttitle{Boulders of Dimorphos}
\shortauthors{Jewitt}

\begin{document}

\title{The Dimorphos Boulder Swarm }


\author{David Jewitt$^{1}$, Yoonyoung Kim$^2$, Jing Li$^1$ and Max Mutchler$^3$ \\
} 
\affil{$^1$Department of Earth, Planetary and Space Sciences,
UCLA}
\affil{$^2$Lunar and Planetary Laboratory, University of Arizona}
\affil{$^3$ Space Telescope Science Institute, 3700 San Martin Drive, Baltimore, MD 21218}

\email{jewitt@ucla.edu}

\begin{abstract}
We present deep Hubble Space Telescope images taken to examine the ejecta from the DART spacecraft impact into asteroid Dimorphos. The images reveal an extensive  population of co-moving boulders, the largest of which is $\sim$7 m in diameter (geometric albedo 0.15 assumed).  Measurements of 37 boulders show a mean sky-plane velocity dispersion of 0.30$\pm$0.03 m s$^{-1}$, only slightly larger than the 0.24 m s$^{-1}$ gravitational escape velocity from the Didymos/Dimorphos binary system.  The total boulder mass, $M_b \sim 5\times10^6$ kg (density 2200 kg m$^{-3}$ assumed), corresponds to about 0.1\% of  the  mass of Dimorphos and the boulders collectively carry about 3$\times10^{-5}$ of the kinetic energy delivered by the DART spacecraft impact. The sky-plane distribution of the boulders is asymmetric, consistent with impact into an inhomogeneous, likely rubble-pile, body.  Surface boulder counts on Didymos show that the observed boulder swarm could be ejected from as little as 2\% of the surface of Dimorphos (for example a circular crater at the impact point about 50 m in diameter).  The large, slow-moving boulders are potential targets to be investigated in-situ by the upcoming ESA HERA mission.
\end{abstract}

\keywords{asteroids: general---asteroids: individual Dimorphos }

\section{INTRODUCTION}
\label{intro}

Near-Earth asteroid  65803 Didymos and its diminutive companion Dimorphos form a compact binary system the basic parameters of which are known from radar observations (Naidu et al.~2020) and from a series of mutual occultation events (Pravec et al.~2022).  The two components are about 800 m and 160 m in diameter, respectively, and separated by only $\sim$1.2 km, with an orbit period of $\sim$11.9 hours.   Didymos rotates rapidly with a period $P$ = 2.26 hour, while the smaller body is presumed to be in synchronous rotation with the orbit. The density of Didymos is $\rho$ = 2200$\pm$400 kg m$^{-3}$ (Naidu et al.~2020); the density of Dimorphos is presumed to be the same.   Taken together, the known physical properties suggest that Dimorphos formed by the accumulation of debris released from Didymos as a result of past rotational instability.  Other parameters of the Didymos/Dimorphos system are conveniently tabulated by Rivkin et al.~(2021).

Dimorphos was impacted by the NASA DART spacecraft on UT 2022 September 26, resulting in the ejection of debris and the formation of a long, comet-like tail swept in the antisolar direction by solar radiation pressure (Li et al.~2023, Graykowski et al.~2023).    The impact is scientifically interesting both as a way to study the mechanical response of a rubble pile to an energetic collision and as a well-characterized analog of certain active asteroids in which mass loss has been shown to result from impact (Jewitt et al.~2010, Kim et al.~2017).  Here, we present  deep post-impact observations of Didymos/Dimorphos with the Hubble Space Telescope  revealing a previously undetected population of large boulders and discuss their nature and origin.

%

\section{OBSERVATIONS}

We  used the 2.4 m diameter Hubble Space Telescope (HST) to observe the Dimorphos debris trail  allocated under programs GO 17289 (4 orbits), 17293 (16 orbits) and 17297 (15 orbits).  All images were taken using the WFC3 camera, which houses two 2015$\times$4096 pixel charge coupled devices (CCDs) separated by a 1.2\arcsec~wide gap.  To reduce the readout time, we utilized only one of the two CCDs, providing an 80\arcsec$\times$160\arcsec~field of view at  0.04\arcsec~pixel$^{-1}$ image scale.  We used the F350 LP filter in order to maximize throughput.  This filter has an effective central wavelength $\lambda_c$ = 6230\AA~when observing a Sun-like (G2V) source and a full-width at half maximum (FWHM) $\Delta \lambda$ = 4758\AA.

Images from HST suffer from large numbers of cosmic rays, as well as from field contamination by background stars and galaxies that are rapidly swept through the field of view by parallax, which reached peak rates $\sim$220\arcsec~hour$^{-1}$ (1.5 pixel s$^{-1}$) during the observations. Field stars and galaxies were trailed by up to 11\arcsec~(300 pixels) in each image.   The images were dithered in order to provide protection from defective CCD pixels, and we made no attempt to control the spacecraft orientation angle so that the direction to astronomical north relative to the edge of the CCD is variable.  As a result, the position and orientation of the Dimorphos trail on the CCD array both change from image to image.   The geometrical circumstances of observation are given in Table \ref{geometry}.

\section{RESULTS}
\subsection{Images}

\begin{figure}
\epsscale{0.95}
\plotone{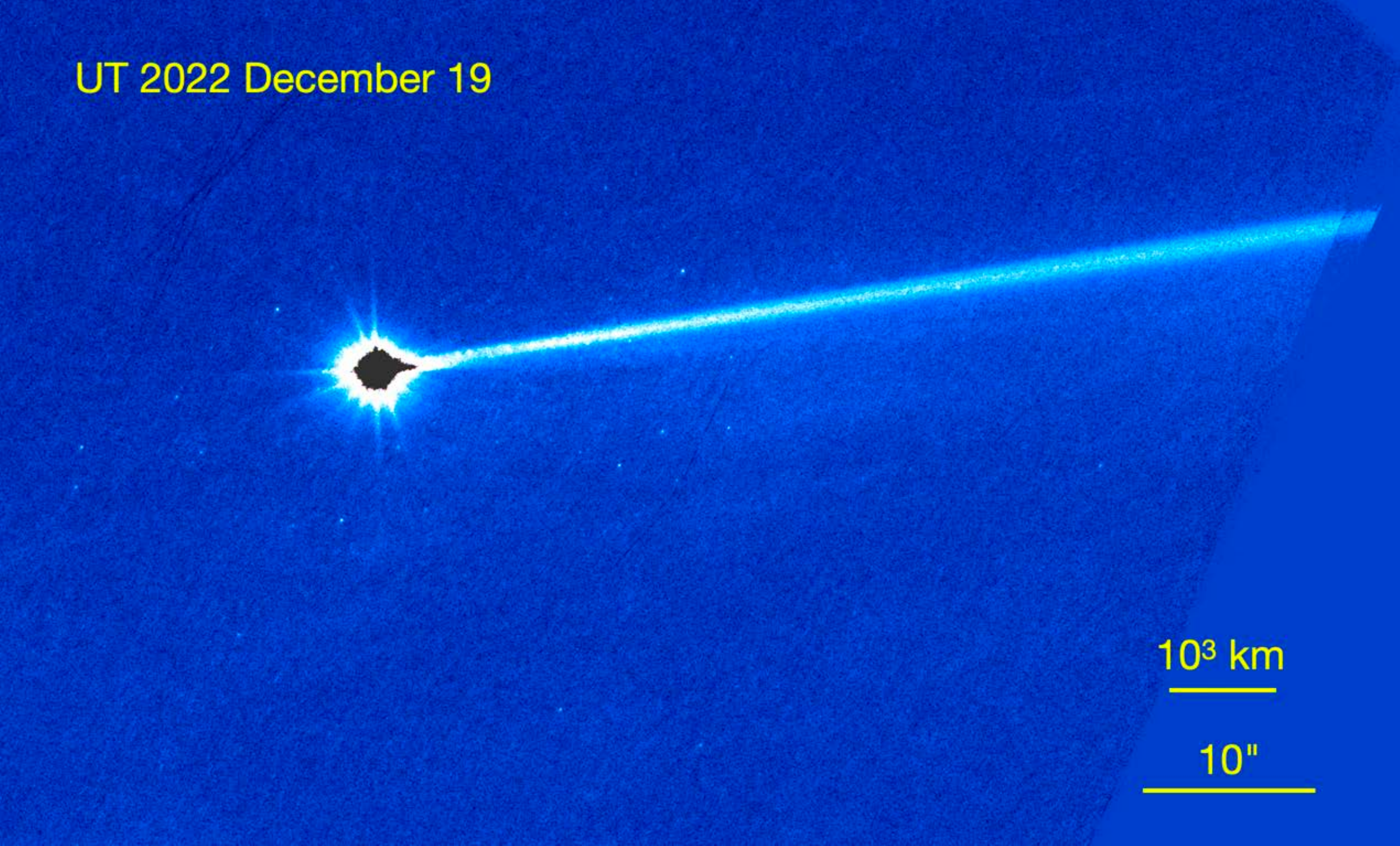}
\plotone{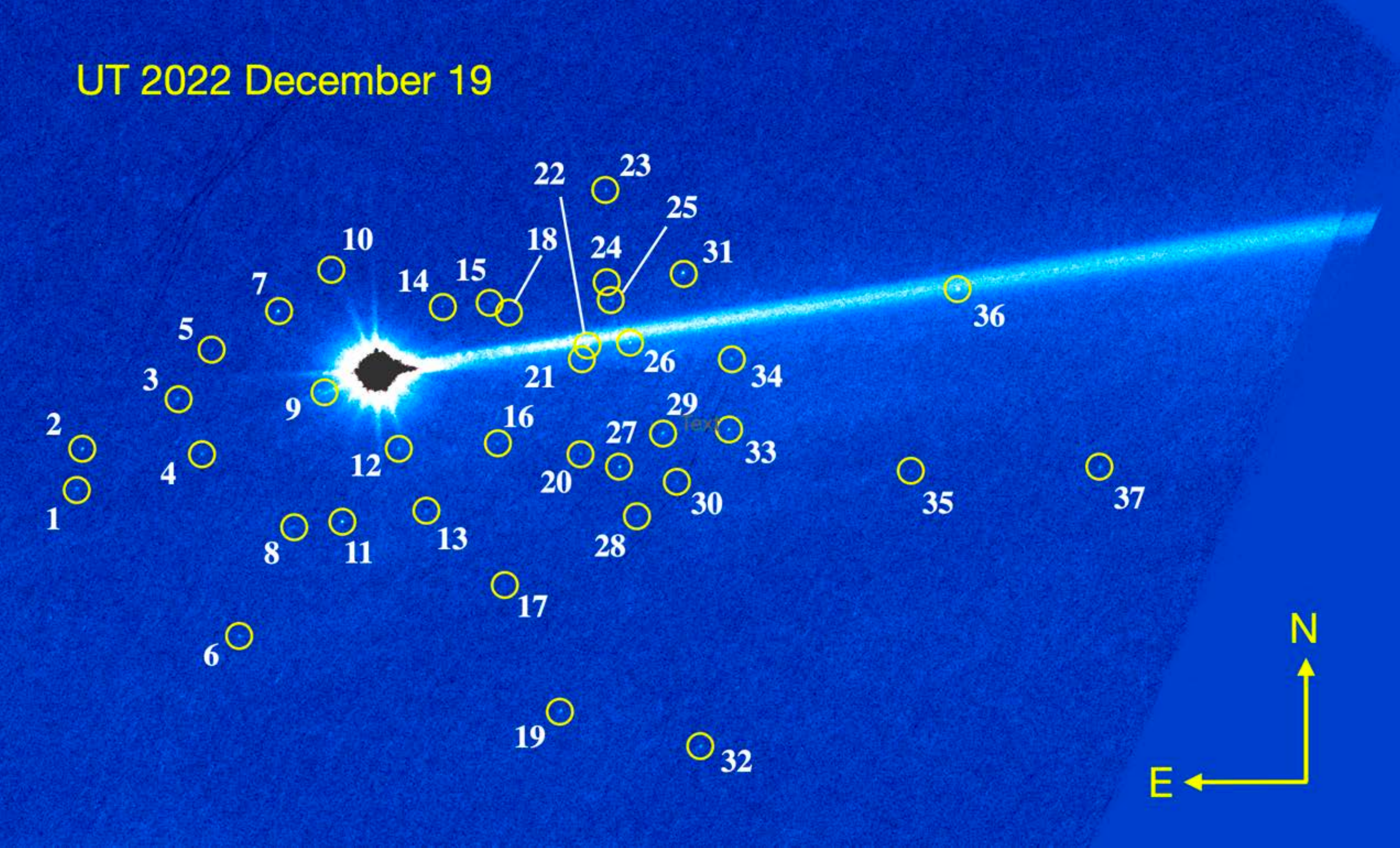}
\caption{(Upper:) Wide field composite image of the Dimorphos trail. The diagonal line to the right marks the edge of the CCD field of view.  (Lower:) Same with individual co-moving sources circled and numbered.   Scale bars show 1000 km and 10\arcsec.  Image has North to the top, East to the left. \label{wide}}
\end{figure}

We  first rejected images in which the target was lost or trailed owing to guide star problems with the HST.  We then shifted the images to a common center and rotated them to bring North to the top.  Both cosmic rays and trailed background object contamination were suppressed by computing the medians of image subsets.  Figure \ref{wide} shows the December 19 image composite constructed from all 24 images.  The multiple spikes on the (saturated) image of Didymos/Dimorphos are the telescope diffraction spikes and CCD charge-transfer trails of individual images rotated to bring the field of view into alignment.  The most prominent feature of the image is the debris trail (c.f.~Li et al.~2023, Graykowski et al.~2023) which extends between the projected antisolar and projected orbit directions and which is caused by the action of solar radiation pressure on centimeter particles.  The 2022 December 19 composite also  shows a set of point sources co-moving with Dimorphos.   The point sources cannot be CCD defects because they share the motion of Didymos/Dimorphos in images that are dithered and also rotated by different angles.   To test the possibility that the co-moving objects might be residual noise clumps in the sky background, we divided the 24 images into two groups of 12 and from them computed separate image composites. When compared, different subsets of the data  consistently revealed the  co-moving objects, all of which  maintained positions fixed with respect to Dimorphos even as the pointing and orientation of the WFC3 field change.  Once noticed in image composites, the brighter objects are evident even in individual images from the 24 image sequence on December 19. Moreover, their point-spread functions (PSFs) are consistent with the $\sim$0.08\arcsec~PSF of WFC3.  Consequently, we interpret the co-moving sources as a population of boulders ejected from Didymos/Dimorphos by the spacecraft impact.  

\begin{figure}
\epsscale{0.99}
\plotone{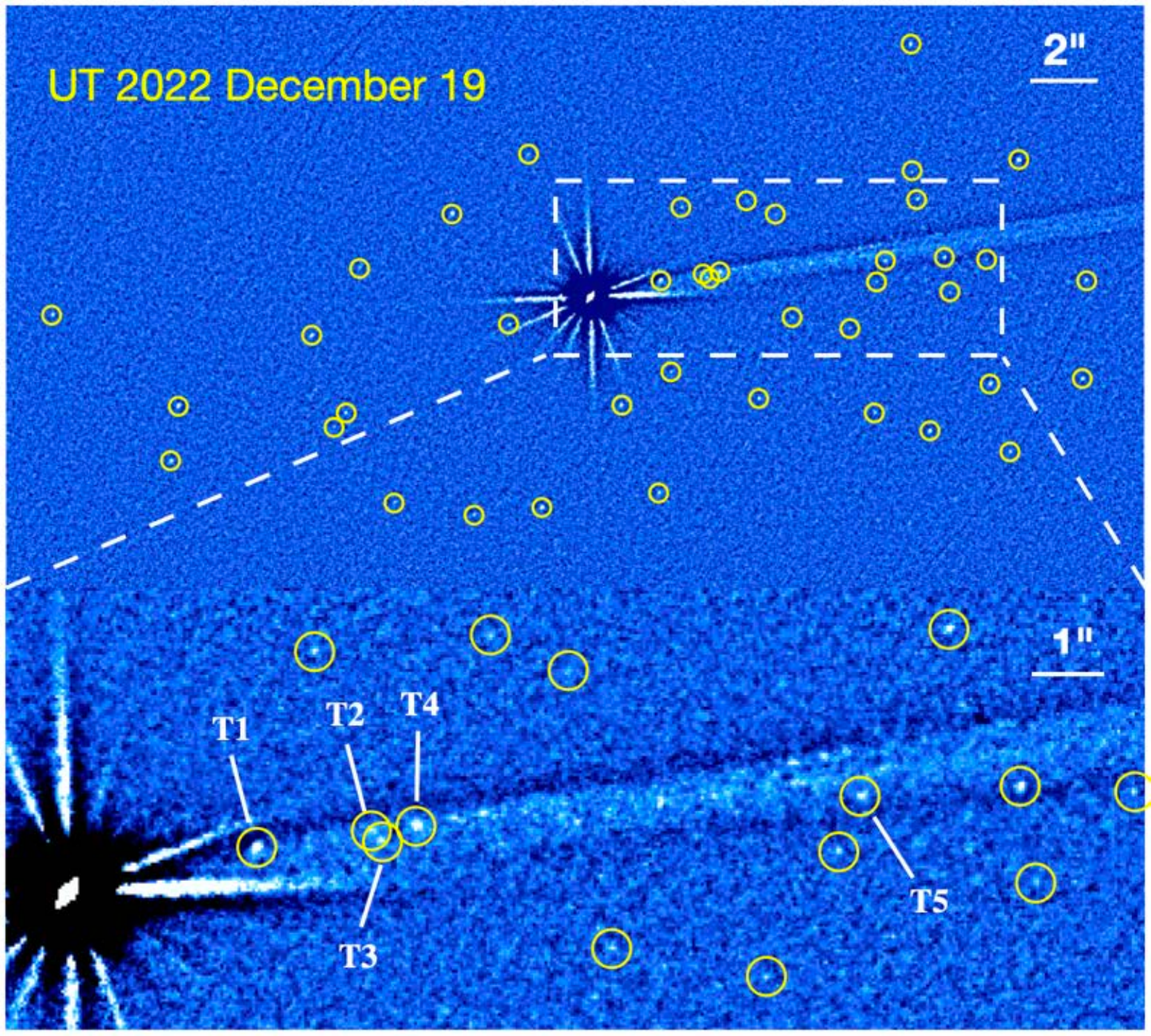}
\caption{(Upper:) Magnified and nucleus-centered view of Figure \ref{wide} shown spatially filtered to suppress diffuse emission, and revealing co-moving sources embedded in the dust trail.   The dashed box has dimensions 400 km x 1000 km. (Lower:) The portion of the upper image within the dashed box shown magnified with objects in the trail labeled.  Scale bars are given in the upper right of each panel. Images have North to the top, East to the left.\label{smoothed}}
\end{figure}

Thirty-seven boulders are circled and numbered in the lower panel of Figure \ref{wide}.    The sky-plane positions of the boulders are given in Table \ref{positions}.  Objects affected by scattered light or field objects are marked ($\star$) in the Table and excluded from analysis here.   The fainter co-moving objects are difficult to see at the scale of Figure \ref{wide}; a magnified portion of this image is shown in Figure \ref{smoothed} where, in addition, the image has been spatially filtered to suppress diffuse light.  Filtering was done by self-subtracting an image convolved with a Gaussian kernel having a full width at half maximum 9.4 pixels (0.38\arcsec).  The suppression of the debris trail reveals objects embedded in and near the diffuse trail, labeled T1 through T5 in Figure \ref{smoothed}, while the reality of fainter objects in the trail is under investigation.  These more difficult to measure trail objects are not further studied here.

\begin{figure}
\epsscale{0.99}
\plotone{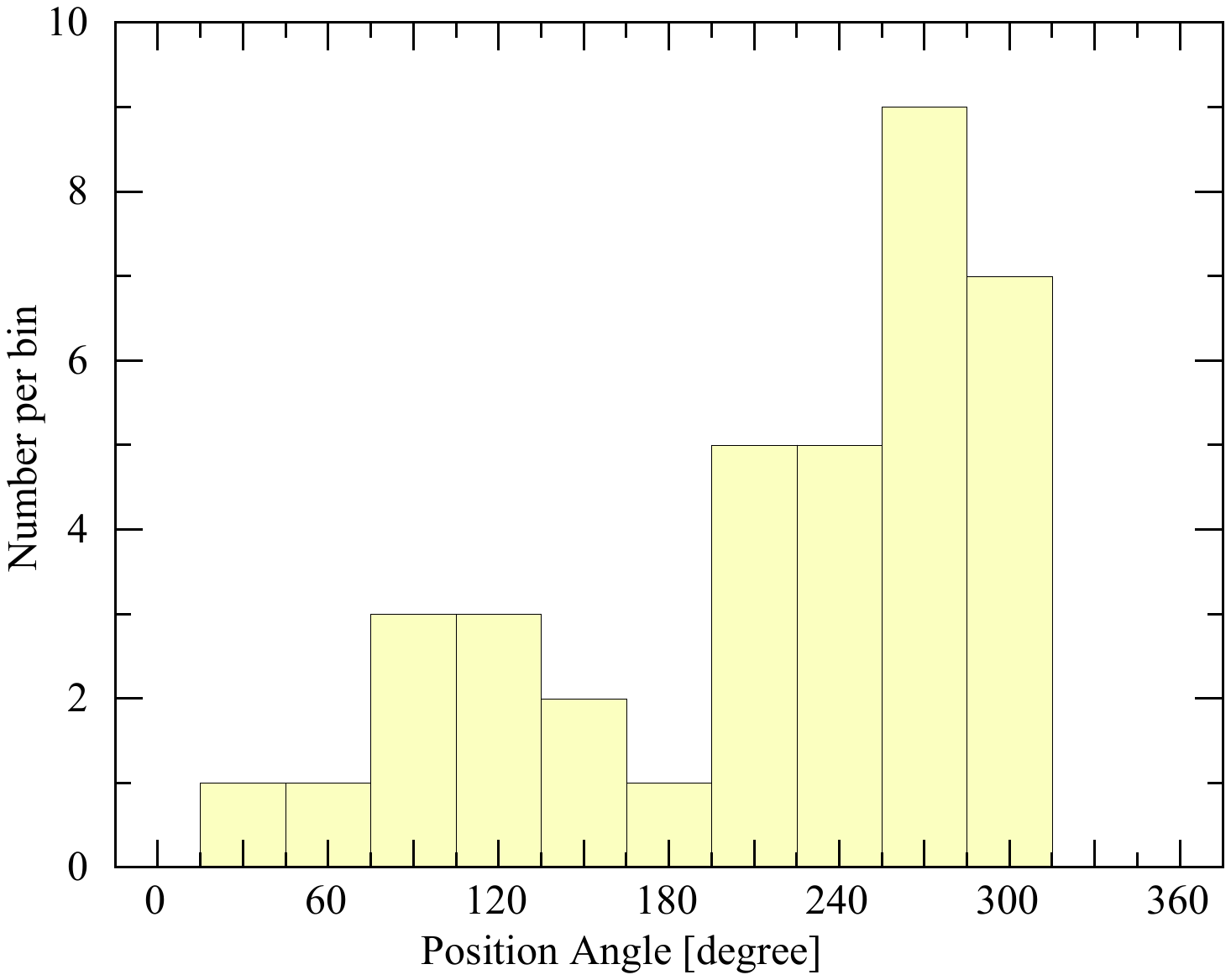}
\caption{Histogram of boulder position angles on UT 2022 December 19 measured with respect to Didymos/Dimorphos showing that the distribution is anisotropic.
\label{position_angles}}
\end{figure}

The boulders are broadly distributed around Didymos/Dimorphos, similar to those in natural impact remnant P/2010 A2 (Jewitt et al.~2010, Kim et al.~2017) and unlike those narrowly aligned in the rotationally unstable active asteroid 331P/Gibbs (Jewitt et al.~2021, Hui and Jewitt 2022).   The angular distribution of the boulders relative to Didymos/Dimorphos is clearly anisotropic.  Figure \ref{position_angles} shows a histogram of the position angles of the boulders relative to the Didymos/Dimorphos binary, where a broad peak along position angle 270\degr~is evident.   Some $\sim$70\% of the boulders are located to the west of Dimorphos and $\sim$80\% are located south of the projected orbit.  

\begin{figure}
\epsscale{0.99}
\plotone{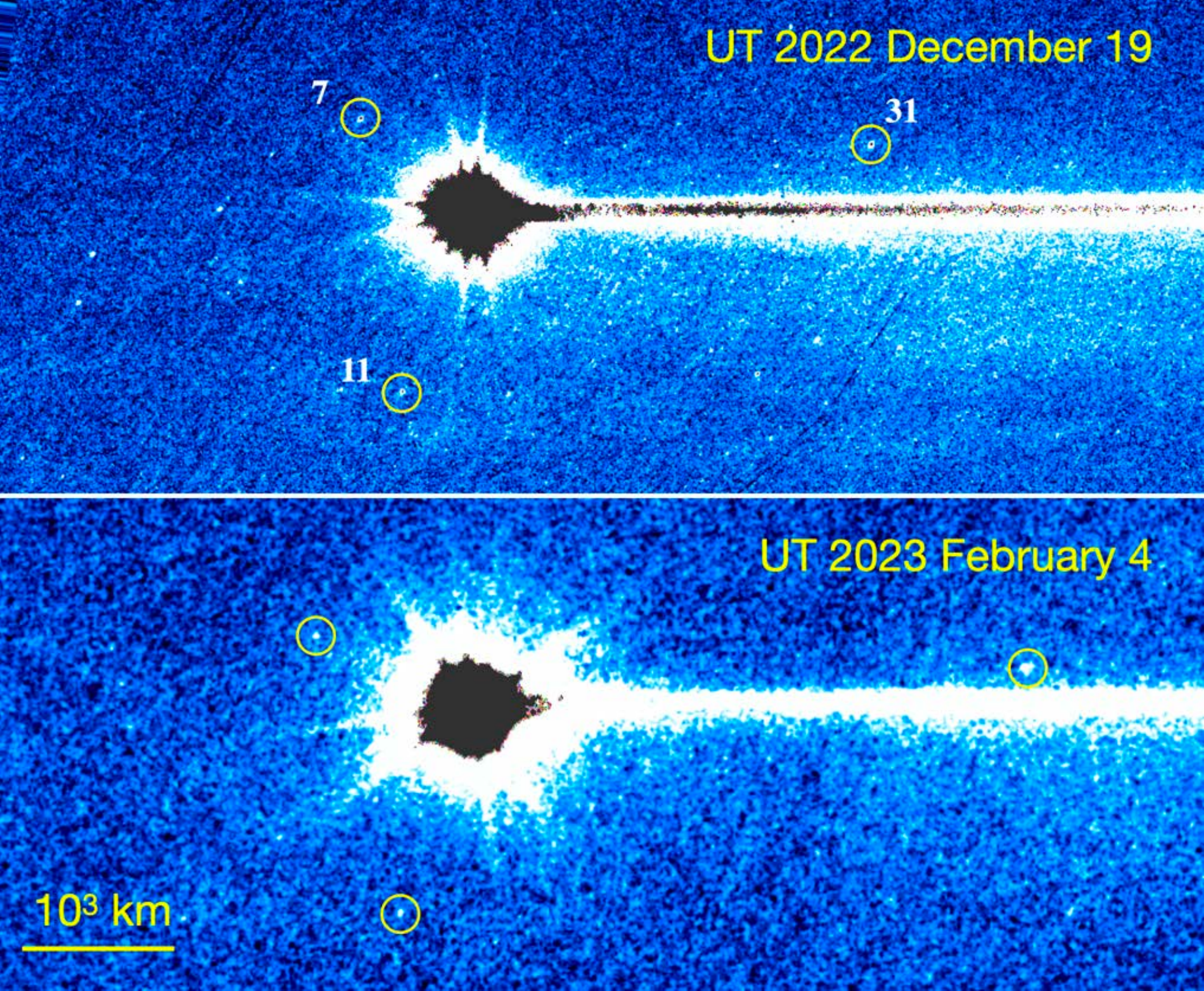}
\caption{Comparison of composite images from UT 2022 December 19 and 2023 February 4.  The images have been rotated to bring the dust trail into horizontal alignment, scaled for the different geocentric distances (c.f.~Table \ref{geometry}) and the displays are heavily stretched to show faint structures, causing bright pixels to appear clipped.  The three brightest boulders are circled in both panels, while others are not marked for clarity.  A 1000 km scale bar is shown. \label{2dates}}
\end{figure}

By comparison, the  composite image from UT 2023 February 4 shows only the brightest boulders owing to the less favorable viewing geometry.  The 2023 February composite is compared with that from 2022 December in Figure \ref{2dates} where both images have been rotated to bring the trail direction to the horizontal, and the image from 2023 February has been scaled by the ratio of the geocentric distances, taken from Table \ref{geometry}.  The 1000 km scale bar in the lower panel applies to both images.  The three brightest are circled together and tenatively linked to their likely counterparts in the 2022 December panel.    

\subsection{Photometry}
\label{sizes}
The following analysis is based mainly on the data from UT 2022 December 19.  Photometry of the brightest boulders is summarized in Table \ref{photometry}.  We used circular  apertures 5 pixels (0.2\arcsec) in radius with background subtraction from the median signal in a contiguous annulus having outer radius 15 pixels (0.6\arcsec).   Substantially larger apertures give larger uncertainties due to the sky background, while smaller apertures are unsatisfactory as they exclude a significant fraction of the light in wings of the PSF, which has a full width at half maximum $\sim$2 pixels.  The Table shows that the brightest boulders have apparent magnitudes $V \sim$26.5 in the 2022 December data.

The difference between the absolute magnitudes, $H$, and the apparent magnitudes, $V$, is a function of the observing geometry (Table \ref{geometry}) and given by  

\begin{equation}
V - H =  2.5 \log_{10}(r_H^2 \Delta^2) + \Phi(\alpha)
\label{absolute}
\end{equation}

\noindent where $r_H$ and $\Delta$ are the heliocentric and geocentric distances in au, respectively, and $\Phi(\alpha)$ is a correction from observation phase angle $\alpha$ to $\alpha$ = 0\degr.  The phase function of the Didymos/Dimorphos binary is not known.  We used the averaged phase function of S-type asteroids from Shevchenko et al.~(2019)   and list the resulting values of $V-H$ for each epoch of observation in Table \ref{geometry}.   The Table shows that, relative to the data from 2022 December 19, the expected fading of point sources is 2.08 magnitudes by 2023 February 4, rising to 4.91 magnitudes by 2023 April 10.  This progressive geometric fading was too strong to be countered by increasing the number of orbits used on each date and this accounts for the decrease in the number of detected boulders with time from the impact.  The April 10 data (Table \ref{geometry}), in particular, were obtained later than planned as a result of practical difficulties with HST scheduling. They suffer so strongly from the rapidly rising geocentric distance that no boulders are reliably detected and we do not show these data here.

The effective diameters of the boulders are related to their absolute magnitudes by

\begin{equation}
D_b = \frac{1.33\times10^6}{p_V^{1/2}}10^{-0.2H},
\end{equation}

\noindent where $p_V$ = 0.15 is the adopted albedo of the Didymos/Dimorphos system (Naidu et al.~2020), which we assume to be the same as the albedo of the boulders.  Here, $D_b$ is the diameter of a circle having the same cross-section as a given boulder, regardless of its shape.  The uncertainties on the derived diameters of the brightest boulders are systematic, and depend on the assumed albedo and the phase function correction.  For example, an error in the phase function correction by $\pm$0.2 magnitude, would lead to  $\pm$10\% errors in the photometrically derived diameters of boulders, which is unimportant for our present purposes. 

 The binned distribution of boulder diameters is shown in Figure \ref{sizes} where a roll-over in the count occurs for $D_b \le$ 4 m.  Smaller boulders are under-counted because their average magnitudes are too faint to be distinguished against the background noise in the data.  We fitted a power law to the 15 boulders with $D_b >$ 4 m, weighted according to Poisson statistics, finding a differential size index $q$ = -3.9$\pm$1.5.  The relatively large uncertainty on $q$ results from the modest number of boulders and the limited size range of measurable boulders, 4 m $\le D_b \le$ 7 m.   Models of the dust trail give $q$ = -3.7$\pm$0.2 for particles in the millimeter to centimeter size range (Li et al.~2023).

\subsection{Lightcurves}
We measured lightcurves of the brighter boulders in the December 19 data, again using photometry in 5 pixel (0.2\arcsec)  radius apertures with background subtraction from the median signal in a contiguous annulus having an outer radius of 15 pixels (0.6\arcsec). The lightcurves span the $\sim$5 hour period during which HST observed the target, with gaps in the coverage corresponding to the $\sim$1.6 hour orbit period of the telescope.  Images in which the photometry was affected by the proximity of a nearby cosmic ray strike, or by background structure due to the passing of a field star or galaxy, were rejected.  

The measurements are limited by the signal-to-noise ratios achieved in individual 193 s integrations (e.g.~targets with V = 26.5, 27.0, 27.5  have single image signal-to-noise ratios 4.4, 2.8, 1.8, respectively). Consequently, rotational periods cannot be reliably determined for most boulders.  The most convincing exception is that of Boulder 31, which shows short-term variations in brightness larger than the  uncertainties in the photometry and a systematic brightening trend (0.073$\pm$0.028 magnitudes per hour) through the five hour observing window (Figure \ref{boulder31}).  We interpret the short-term variations as  modulation of the projected boulder cross-section owing to rotation and irregular shape.  The brightening trend, shown by the solid black line in  Figure  \ref{boulder31}, could indicate that Boulder 31 is in an excited rotational state, with a period much longer than five hours.  Excited rotational states are to be expected following impulsive ejection of the boulder.  We used de-trended data and phase dispersion minimization (PDM) to examine periodicity in the lightcurve.  The results are consistent with a lightcurve period 0.413 hour (0.826 if doubly periodic) and a peak-to-peak range $\Delta V \sim$ 1.0 magnitudes (Figure \ref{boulder31}).  However,  the $\sim$1.6 hour spacing in the temporal coverage imposed by the orbital period of HST introduces aliasing of the data and the period is not a unique solution for the lightcurve.  The  rotation period is short compared to the $\sim$11.9 hour rotation period of Didymos and to the $\sim$2.2 hour rotational barrier observed in kilometer sized and larger asteroids (e.g.~Pravec et al.~2002).  However, the typical period of near-Earth asteroids smaller than 60 m, themselves presumably products of asteroid breakup, is $\sim$0.7 hours (Hatch and Wiegert 2015) so that Boulder 31 is not in this regard unusual. While the boulder must be in a state of internal tension because of  rapid rotation, this can be counteracted by even very modest material strength.  We calculate, for example, that Boulder 31 could be held together by a cohesive strength as small as 0.1 N m$^{-2}$,  no stronger than a clod of dirt, in order to resist rotational disruption.

Rotational periods of the other boulders could not be usefully defined given their faintness and the presence of many aliases.  Even without determining the periods, we used the lightcurve range, $\Delta V$, for the brighter asteroids to estimate the projected boulder shapes from $a/b = 10^{0.4\Delta V}$.   For the five brightest boulders in Table \ref{photometry} we find a mean $\overline{\Delta V} = 0.34\pm0.18$ magnitudes, corresponding to $a/b \sim 1.4\pm0.2$.  This can be compared to the axis ratios of small near-Earth asteroids ($a/b$ = 1.4; Hatch and Wiegert 2015) and to boulders found on asteroids Eros, Ryugu and Itokawa ($a/b$ = 1.4, 1.5 and 1.6, respectively; Michikami and Hagermann.~2021).

\subsection{Ejection Speeds}
We crudely estimate the projected speeds of the boulders  from their spatial distribution in the December data.  Specifically, in the data from UT 2022 December 19, the mean and median angular separations of the boulders from Didymos/Dimorphos are $\overline{\Delta \ell}$ = 13.7\arcsec$\pm$1.2\arcsec~and $\Delta \ell_M$ = 12.7\arcsec, respectively, corresponding to 2175$\pm$190 km and 2016 km in the plane of the sky.  Assuming an origin at the time of the DART impact $\Delta t$ = 83 days (7.17$\times10^6$ s) before the December 19 image, the average and median speeds in the plane of the sky are $\Delta V = \overline{\Delta \ell}/\Delta t = 0.30\pm0.03$ m s$^{-1}$ and $\Delta \ell_M/\Delta t$ = 0.28 m s$^{-1}$, respectively.  There is a significant dispersion in the projected boulder speeds: the slowest 5\%  of the boulders have projected speeds $\Delta V \le$ 0.10 m s$^{-1}$ while the fastest  5\% have  $\Delta V \ge$  0.67  m s$^{-1}$.   We searched for a correlation between the size of the ejected boulders (inferred from photometry) and their ejection speed (inferred from the distance traveled), finding none.  The absence of a size vs.~speed trend is consistent with a model of impacted asteroid P/2010 A2 (Kim et al.~2017) and with laboratory experiments showing only a weak dependence of speed on size (Nakamura and Fujiwara 1991, Nakamura et al.~1992, Raducan et al.~2022).

We seek to interpret the boulder speeds. First, we  note that the Hill sphere of the Didymos/Dimorphos system at the perihelion distance is $\sim$130 km in diameter, corresponding to about 0.8\arcsec~(20 HST pixels) in the 2022 December data.  Therefore, the detected boulders all lie outside the Hill sphere and can be assumed to follow heliocentric orbits.

 The acceleration due to radiation pressure acting on a perfectly absorbing spherical object of diameter $D_b$  [m] can be estimated from

\begin{equation}
\alpha = \frac{6 L_{\odot}}{16 \pi \rho c r_H^2 D_b}
\label{diameter}
\end{equation}

\noindent where $L_{\odot} = 4\times10^{26}$ W is the luminosity of the Sun, $c = 3\times10^8$ m s$^{-1}$ is the speed of light, $\rho$ is the density (we assume 2200 kg m$^{-3}$), and $r_H$ is the heliocentric distance in meters.  Substituting, we obtain $\alpha = 3\times10^{-9}$ m s$^{-2}$ for a $D_b$ = 1 meter  boulder at 1 AU.  In $\Delta t$ = 83 days the motion induced by radiation pressure on a 1 m boulder would be only $\alpha \Delta t^2/2 \sim$ 80 km, falling to $\sim$11 km for a $D_b$ = 7 m boulder.  These distances, corresponding to 0.5\arcsec~and 0.07\arcsec, respectively, are within the saturated image core of Figure \ref{wide} and are  very small compared to the $\sim10^4$ km boulder separations in the figure.   Therefore, the motions of the boulders between ejection and the December data can be treated as unaffected by radiation pressure.  In this case, $\Delta V = 0.30\pm0.02$ m s$^{-1}$  gives a true measure of the average boulder ejection speed projected into the plane of the sky.

Boulders  must be launched faster than $\sim$0.09 m s$^{-1}$ in order to escape from Dimorphos but faster than $\sim$0.24 m s$^{-1}$, the escape speed from the Didymos/Dimorphos pair, in order to escape the system into interplanetary space.  DART ejecta with launch speeds $\Delta V <$ 0.09 m s$^{-1}$ should fall back to the surface of Dimorphos on a timescale of hours.  Ejecta with speeds 0.09 $\le \Delta V \le$ 0.24 m s$^{-1}$ should be trapped in temporary orbit about the system barycenter, leading eventually to impact with one or other component of the binary or to acceleration via scattering in near-miss encounters producing delayed escape (timescales from the half-day crossing time to months) from the system.

The average boulder speeds are comparable to the Didymos/Dimorphos system escape speed, $V_{DD}$ = 0.24 m s$^{-1}$, showing that the boulders are amongst the slowest of the ejected bodies that were able to escape the system.  Strictly, $\Delta V$ is a measure of the excess speed after a boulder has climbed out of the potential well of the asteroid binary. The launch speed, $U$, is related to $\Delta V$ by $U = (\Delta V^2 + V_{DD}^2)^{1/2}$.   By this relation, the measured average speed $\Delta V$ = 0.30 m s$^{-1}$  corresponds to a mean launch speed $U$ = 0.38 m s$^{-1}$, while 5\% of the detected boulders were launched with $U \le$ 0.26 m s$^{-1}$ and 5\% with $U \ge$ 0.71 m s$^{-1}$.

\subsection{Mass and Energy}

The sum of the spherical-equivalent volumes of the boulders listed in Table \ref{photometry} is $\sim$1206 m$^3$ and, given density $\rho$ = 2200 kg m$^{-3}$, their combined mass is 2.6$\times10^6$ kg.  Assuming a differential power law distribution of boulder sizes like that found on the surface of Dimorphos (see Appendix),  boulders in the 4 m to 7 m size range of Table \ref{photometry} contain about 50\% of the mass in the entire distribution, giving a total ejected mass estimate of 5.2$\times10^6$ kg.  This is about 0.1\% of the $4\times10^9$ kg mass of Dimorphos and about 10$^4$ times the mass of the DART impactor.   The ejected mass is comparable to models of impact  into a dense-packed  field of 7 m sized boulders (Panel (c) from Raducan et al.~2022) and of the same order as the nominal 1.5$\times10^6$ kg mass loss predicted by Fahnestock et al.~(2022).  A slightly larger mass, 0.3\% to 0.5\%~of the mass of Didymos, was inferred from the diffuse material (Graykowski et al.~2023) while a wider range, 0.2\% to 1.2\%,  was determined from submillimeter wavelength observations (Roth et al.~2023).  Since the boulders are traveling very slowly, they carry only a small fraction ($\sim$10\%) of the recoil-dominated momentum delivered to Dimorphos by the impact and their contribution to the deflection of Dimorphos awaits consideration of the (currently unknown) angular distribution of the boulder ejection velocities.

The total kinetic energy of  boulders having a combined mass $M_b$ = 5$\times10^6$ kg  and launched at the mean  speed $U$ = 0.38 m s$^{-1}$ is $E_b = 3.6\times10^5$ J.   The mass (500 kg) and speed (6.6 km s$^{-1}$) of the DART spacecraft give an impact energy $E_D = 1.1\times10^{10}$ J.   The ratio, $\varepsilon = E_b/E_D = 3\times10^{-5}$, shows that very little energy was converted into motion of the boulders.

\section{Discussion}
\subsection{Boulders on the Surface of Dimorphos}
Dimorphos is an ellipsoidal body with estimated axes 
177 m$\times$174 m$\times$160 m (Daly et al.~2023) and total surface area approximately $A = 7.6\times10^4$ m$^2$, or about 0.08 km$^2$.   
One hemisphere of Dimorphos was  imaged in full from the approaching DART spacecraft, but only a portion of the surface in the vicinity of the impact point was imaged at high resolution.  We used the penultimate image obtained using the DRACO camera (Fletcher et al.~2022) on the DART spacecraft (Figure \ref{penultimate}) to study the boulders on the pre-impact surface of Dimorphos.  This image, recorded from a range of 12 km, shows a boulder-strewn region  with dimensions approximately 30 m $\times$ 30 m, corresponding to $\sim$1\% of the total surface area of Dimorphos.  Meter-sized and larger boulders  are well-resolved in shape and texture, while features smaller than $\sim$0.1 m can be discerned.   The outlines of many sub-meter rocks are confused by the complexity of the scene, being partially hidden by shadows and obscured by overlapping boulders.  Moreover, in practice it is difficult to distinguish the outline of a small boulder from a sun-catching, topographically high region on the surface of a larger boulder.  However, these problems are less important for meter-sized and larger boulders to which, for these reasons, we confine our analysis.  The impact location of the DART spacecraft according to Daly et al.~(2023) is marked in Figure \ref{penultimate} by a yellow circle.

We projected Figure \ref{penultimate} on a screen and measured the  longest, $2a$, and shortest, $2b$, dimensions of each clearly discerned boulder, scaled to the 30 m width of the field of view.  Represented  as an ellipse, the boulder cross-section is simply $\pi ab$.  We define the effective boulder diameter, $D_b$,  as the diameter of a circle having the same area as the boulder, $\pi D_b^2/4 = \pi ab$, or  $D_b = 2 (ab)^{1/2}$.    
By this measure the largest boulder, Atabaque, which is centrally located in Figure \ref{penultimate}, has $D_b \sim$ 5 m and so is comparable in scale to the larger objects detected in HST data.

We  used the boulder counting data to  measure the  size distribution on the surface of Didymos. Figure \ref{boulders} shows the numbers of boulders counted within bins 0.2 m wide.  As noted above, small boulders could not be reliably measured due to overlap between boulders,  shadowing effects and image resolution.   These effects are responsible for the roll-over in Figure \ref{boulders} at $D_b \lesssim$ 1 m.  We fitted a differential power-law only to boulders with $D_b \ge$ 1 m by weighted least-squares, assuming Poisson statistics (i.e.~the uncertainty in the count within each bin is equal to the square root of the number of boulders in the bin).  The resulting best fit for the differential power law distribution, scaled to the full 0.08 km$^2$ surface area of Dimorphos,  gives

\begin{equation}
N(D_b) dD_b = (14\pm1) \times10^3 D_b^{-2.8\pm0.4} dD_b.
\label{bouldersize}
\end{equation}

\noindent  By this relation, the number of boulders on the surface with $D_b \ge$ 4 m is $N(D_b \ge 4)$ = 660, compared with 15 boulders of this diameter or larger detected from HST (as listed in Table \ref{photometry}).  The  HST-detected boulders  thus constitute a fraction $f \sim$ 2\% of the number present on the whole surface of Dimorphos.   This is consistent with the ejection of all the boulders from an area $Af \sim$ 1800 m$^2$ of the surface, equal to a circular patch around the impact site with a diameter $2(Af/\pi)^{1/2} \sim$ 50 m, or with partial ejection from a larger area.  For comparison, the DART Impact Modeling and Simulation Working Group tentatively estimate a minimum crater diameter 40 m to 60 m (Stickle et al.~2023).

At first sight, the boulder size distribution is surprisingly flat.  However, we note that the index derived for  Dimorphos  is  similar to that measured locally for meter scale boulders on sub-kilometer rubble pile objects Ryugu and Bennu (Schroder et al.~2021).  Figure \ref{michi} shows  that the power law distribution becomes steeper as the boulder size increases. The physical reason for the systematic trend towards steeper power laws on larger objects is unclear.  We note that the larger boulders with the steeper distribution in the Figure are found on the surfaces of larger, higher escape velocity bodies like Vesta, Ceres and the Moon. This might suggest that the flatter distributions on small rubble pile objects including Ryuga, Bennu and Dimorphos are the result of the selective escape of smaller, faster impact-produced fragments, whereas escape from Vesta and Ceres sized asteroids is negligible.    As noted in Section \ref{sizes}, the size distribution of the ejected boulders detected in HST data is matched by $q$ = -3.9$\pm$1.5, which is within 1$\sigma$ of the index in Equation \ref{bouldersize}.

\subsection{Ejection Mechanisms}

We briefly consider two distinct origins for the boulders assuming that they were released in response to the DART impact.   First, the boulders could be directly ejected from the impact site as part of the crater-forming process.  Second, the boulders could be pre-existing objects launched from the surface of Dimorphos by seismic shaking.   We note that these two processes are not mutually exclusive, with boulders being ejected both in the impact cone and more broadly by seismic shaking.  A third possibility has been suggested, namely that  mass loss from the parent body, Didymos, could be triggered by the impact of ejecta from the DART impact on Dimorphos  (Hirabayashi et al.~2022). We do not consider this possibility here because we do not know how to make a connection to observable quantities.

\textbf{Cratering:} Hypervelocity impacts eject material in a hollow cone configuration, with the axis of the cone roughly anti-parallel to the impact direction. Particles ejected in such a cone at a given speed define a ring, expanding in proportion to the time since ejection.  When viewed  90\degr~from the axis and spacecraft direction (as at the time of impact in 2022 September), the edges of the cone appear as two ``wings'' separated by a large angle (e.g.~panel c of Figure 3 in Li et al.~(2023), where this angle is reported to be 125\degr$\pm$10\degr).  (We note that the empirical cone axis differed from the impact direction by $\sim$21\degr (Cheng et al.~2023) but this difference does not materially affect the discussion here). Viewed along the axis of the cone, the ring would appear as a circular annulus while for any other viewing direction, it would appear as an ellipse.  We simulated this ejection cone geometry for the known impact trajectory of DART and for the observational circumstances on 2022 December 19, when the line of sight was inclined to the impact direction by 11\degr.   As expected, the assumption of a uniformly filled ejection cone produces a symmetric distribution of boulders in the plane of the sky, inconsistent with the empirical distribution shown in Figure \ref{position_angles}.   To match the observed non-uniform distribution of boulder position angles requires an ejection cone in which the number density of boulders varies with the azimuth angle around the cone, with more ejected to the South and West than to the North and East.  Impact into an inhomogeneous target surface, for example one dotted with boulders larger than the DART impactor (c.f.~Figure \ref{penultimate}), should produce asymmetries in the ejecta cone (Ormo et al.~2022), as might the curvature of the surface of Dimorphos.  Future measurements of the boulders might allow a better determination of the distribution of ejection velocity vectors.

\textbf{Seismic Shaking:} A second possibility is that internally propagated impact shocks reflect from and accelerate the surface of the asteroid, launching some boulders above the escape speed (Tancredi et al.~2022).  In this case, the directions of boulder launch would be more widely distributed than those of the impact cone.    Although we cannot observationally distinguish seismic shaking ejection from cratering  using the existing data, the former seems less likely given that energy propagation through a rubble pile should be heavily damped (Sanchez et al.~2022).  We note, however, that if the boulders were ejected by seismic shaking, then the energy ratio  $\varepsilon = 3\times10^{-5}$ sets a lower limit to the ``seismic efficiency''.   This quantity has been measured for high speed impacts into different targets over a broad range of energies. It occupies a wide range of values  $5\times10^{-7} \lesssim \varepsilon \lesssim 10^{-3}$ (W{\'o}jcicka et al.~2020).  The collision most similar to that of DART is  the $3\times10^9$ J impact  of the Apollo 12 Lunar Module ascent stage into the lunar regolith (Latham et al.~1970), which produced $\varepsilon \sim 5\times10^{-7}$, about 60 times smaller than here.  Larger values of $\varepsilon$ are associated with impacts involving dense, non-porous solid target material.   The middling value inferred in Dimorphos suggests impact into a medium intermediate between the strongly pulverized regolith of the Moon and solid rock.

Whatever the mechanism of ejection, the trajectories of boulders lifted from the surface of Dimorphos could also have been affected by gravitational scattering from Didymos.  As seen from Dimorphos, Didymos subtends about four steradians and is capable of intercepting and deflecting a fraction of the impact ejecta.  Asymmetry induced by Didymos is a compounding factor in affecting the observed angular distribution of the ejecta.

\subsection{Other Observations}
No ground-based observations have been reported to show even the largest and brightest of the Dimorphos boulders. The independent HST dataset  GO 16674 (PI Jian-Yang Li) employed short integrations and single-orbit visits to the Dimorphos field with the intent to avoid image saturation during the brightest phases of the dust trail development after impact.  Consequently, the data from GO 16674 are not optimized for the study of faint boulders.  However, HST will provide one more opportunity to image the Dimorphos boulders when, in July 2024,  the geocentric distance will fall to $\sim$0.6 au. Following this, no more favorable observing geometry will occur until 2040.

Lastly, the European Space Agency HERA mission is planned to rendezvous with the Didymos/Dimorphos system in late 2026 in order to further investigate the aftermath of the DART impact (Michel et al.~2022).  The small velocity dispersion of the Dimorphos boulders opens up the possibility for in-situ investigation by HERA, and for the study of other boulders too small to be detected even in HST data.   In the more distant future, the release of very low velocity debris by high speed impact suggests the possibility for in-situ sampling of asteroidal material without the need for near-surface operations or landing.


\section{SUMMARY}
Deep images taken with the Hubble Space Telescope reveal a population of slow-moving, meter-sized and larger boulders co-moving with the DART impact target asteroid Dimorphos.  

\begin{itemize}
\item We identify $\sim$40 boulders co-moving with Dimorphos and spread non-uniformly over a region $\sim10^4$ km in extent. The largest boulders have an effective diameter of 7 m while 15 have diameters larger than 4 m (albedo 0.15 assumed).  
\item The combined boulder mass is  5$\times10^6$ kg, equal to about 0.1\% of the 4$\times10^{9}$ kg mass of Dimorphos.
\item The  projected mean speed of the boulders relative to Dimorphos,  0.30$\pm$0.03 m s$^{-1}$, is comparable to the 0.24 m s$^{-1}$ gravitational escape speed from the Didymos/Dimorphos system.  They carry about $3\times10^{-5}$ of the kinetic energy delivered by the DART impactor.
\item The numbers, sizes and shapes of the boulders imaged using HST are consistent with an origin as pre-existing objects dislodged from about 2\% of the surface of Didymos by the DART impact, corresponding to a circular patch 50 m in diameter or larger.
\end{itemize}

\acknowledgments
We thank Jian-Yang Li, Pedro Lacerda, Maarten Roos and the anonymous referee for comments on the manuscript.   Based on observations made with the NASA/ESA Hubble Space Telescope, obtained from the data archive at the Space Telescope Science Institute. STScI is operated by the Association of Universities for Research in Astronomy, Inc. under NASA contract NAS 5-26555.  Support for this work was provided by NASA through grant numbers GO-17289, GO-17293 and GO-17297 from the Space Telescope Science Institute, which is operated by auRA, Inc., under NASA contract NAS 5-26555.



{\it Facilities:}  \facility{HST}.

\clearpage

%

%


\begin{deluxetable}{lccrrrrccrr}
\tabletypesize{\scriptsize}
\tablecaption{Observing Geometry 
\label{geometry}}
\tablewidth{0pt}
\tablehead{\colhead{UT Date \& Time}   &  \colhead{DOY\tablenotemark{a}} &  \colhead{$\nu$\tablenotemark{b}} & \colhead{$r_H$\tablenotemark{c}}  & \colhead{$\Delta$\tablenotemark{d}} & \colhead{$\alpha$\tablenotemark{e}}  & \colhead{$\theta_{- \odot}$\tablenotemark{f}} & \colhead{$\theta_{-V}$\tablenotemark{g}}  & \colhead{$\delta_{\oplus}$\tablenotemark{h}}  & \colhead{Scale\tablenotemark{i}}   & \colhead{V-H\tablenotemark{j}}  }

\startdata
2022 Sep 26 23:14 (Impact)   &  270 & 332.4 & 1.046 & 0.076 & 53.2 & 297.9 & 228.1 & 47.6 & 2.2 & -3.85\\
2022 Dec 19 15:05-20:25  & 353 & 60.3 & 1.178 & 0.219 & 25.4 & 271.4 & 282.8 & -3.9 &  6.4 & -1.98 \\
2023 Feb 4 13:30 - Feb 5 18:38  &  400 - 401 & 93.3 & 1.433 & 0.496 & 21.0 & 110.1 & 274.2 & -6.1 & 14.4 & +0.10\\
2023 Apr 10 10:52 - Apr 11 22:12  & 466 - 467 & 123.0 & 1.771 & 1.282 & 33.7 & 104.2 & 280.6 & -2.1 & 37.1 & +2.93\\

\enddata

\tablenotetext{a}{Day of Year; 1 = UT 2022 January 1 }
\tablenotetext{b}{True anomaly, in degrees }
\tablenotetext{c}{Heliocentric distance, in au}
\tablenotetext{d}{Geocentric distance, in au}
\tablenotetext{e}{Phase angle, in degrees }
\tablenotetext{f}{Position angle of projected anti-solar direction, in degrees }
\tablenotetext{g}{Position angle of negative heliocentric velocity vector, in degrees }
\tablenotetext{h}{Angle from orbital plane, in degrees}
\tablenotetext{i}{Image scale, km pixel$^{-1}$ }
\tablenotetext{j}{Difference between apparent and absolute magnitudes, from Equation \ref{absolute}, in magnitude }

\end{deluxetable}

\clearpage

\begin{deluxetable}{lccrrrrc}
\tablecaption{Boulder Positions on UT 2022 December 19
\label{positions}}
\tablewidth{0pt}
\tablehead{\colhead{N\tablenotemark{a}} &  \colhead{$\Delta RA$\tablenotemark{b}} &  \colhead{$\Delta \delta$\tablenotemark{c}} & \colhead{$\Delta \ell$\tablenotemark{d}} & \colhead{$\theta$\tablenotemark{e}}      }

\startdata
1		&	15.6	&	-6.1	&	16.7	&	111.3	&		\\
2		&	15.3	&	-4.0	&	15.8	&	104.8	&		\\
3		&	10.4	&	-1.4	&	10.5	&	97.9	&		\\
4		&	9.1	&	-4.3	&	10.1	&	115.3	&		\\
5		&	8.6	&	1.1	&	8.7	&	82.8	&		\\
6		&	7.2	&	-13.8	&	15.5	&	152.5	&		\\
7		&	5.1	&	3.1	&	6.0	&	58.6	&		\\
8		&	4.3	&	-8.1	&	9.2	&	152.0	&		\\
9 ($\star$)	&	3.0	&	-1.0	&	3.2	&	109.1	&	\\
10		&	2.3	&	5.3	&	5.8	&	23.7	&		\\
11		&	1.8	&	-7.8	&	8.0	&	167.1	&		\\
12 ($\star$)	&	-1.2	&	-4.0	&	4.2	&	196.0	&	\\
13		&	-2.6	&	-7.3	&	7.7	&	199.4	&		\\
14		&	-3.4	&	3.4	&	4.8	&	314.7	&		\\
15		&	-5.8	&	3.6	&	6.8	&	301.4	&		\\
16		&	-6.3	&	-3.8	&	7.3	&	238.8	&		\\
17		&	-6.6	&	-11.2	&	13.0	&	210.8	&		\\
18		&	-6.9	&	3.0	&	7.6	&	293.7	&		\\
19		&	-9.6	&	-17.7	&	20.1	&	208.4	&		\\
20		&	-10.6	&	-4.3	&	11.4	&	247.8	&		\\
21 ($\star$)	&	-11.7	&	0.7	&	11.7	&	273.3	&	\\
22 ($\star$)	&	-12.4	&	1.0	&	12.5	&	274.8	&	\\
23		&	-11.9	&	9.4	&	15.2	&	308.3	&		\\
24		&	-12.0	&	4.6	&	12.8	&	291.2	&		\\
25		&	-12.2	&	3.6	&	12.7	&	286.6	&		\\
26 ($\star$)	&	-12.4	&	1.1	&	12.5	&	275.0	&	\\
27		&	-12.6	&	-5.0	&	13.6	&	248.4	&		\\
28		&	-13.6	&	-7.6	&	15.5	&	240.9	&		\\
29		&	-14.9	&	-3.2	&	15.2	&	257.7	&		\\
30		&	-15.6	&	-5.8	&	16.6	&	249.7	&		\\
31		&	-15.9	&	5.1	&	16.7	&	287.8	&		\\
32		&	-16.8	&	-19.5	&	25.8	&	220.8	&		\\
33		&	-18.3	&	-3.0	&	18.6	&	260.6	&		\\
34		&	-18.4	&	0.6	&	18.5	&	272.0	&		\\
35		&	-27.8	&	-5.2	&	28.2	&	259.4	&		\\
36 ($\star$)	&	-30.2	&	4.2	&	30.5	&	278.0	&	\\
37		&	-37.6	&	-5.0	&	38.0	&	262.4	&		\\

\enddata

\tablenotetext{a}{~Boulder number (c.f.~Figure \ref{wide}).  Objects for which the position or photometry are affected by field objects or scattered light are marked ($\star$).}
\tablenotetext{b}{~Right Ascension offset from photocenter, arcsecond (positive = East)}
\tablenotetext{c}{~Declination offset from photocenter, arcsecond (positive = North) }
\tablenotetext{d}{~Angular distance from photocenter, arcsecond}
\tablenotetext{e}{~Position angle relative to photocenter, degree}

\end{deluxetable}


\begin{deluxetable}{lccrrrrccrr}
\tabletypesize{\scriptsize}
\tablecaption{Photometry of the Brightest Boulders
\label{photometry}}
\tablewidth{0pt}
\tablehead{\colhead{Number} &  \colhead{V\tablenotemark{a}} &  \colhead{H\tablenotemark{b}} & \colhead{$D_b$\tablenotemark{c}} & \colhead{$P$\tablenotemark{d}}  & \colhead{$\Delta V$\tablenotemark{e}}  & \colhead{$a/b$\tablenotemark{f}}   }

\startdata
%

31	&	26.41	&	28.39	&	7.2	& 0.83 & 1.0 & 2.5		\\													
7	&	26.55	&	28.53	&	6.8	& -- & 0 & 1	\\
11	&	26.69	&	28.67	&	6.3	& -- & 0 & 1	\\
29	&	26.99	&	28.97	&	5.5	& -- & 0.4 & 1.4	\\
27	&	27.06	&	29.04	&	5.3	& -- & 0.3 & 1.3			\\															
2	&	27.06	&	29.04	&	5.4     		\\
19	&	27.19	&      29.17	&	5.0		\\
3	&	27.30	&	29.27	&	4.8		\\
1	&	27.30	&	29.28	&	4.8				\\															
37	&	27.32	&	29.30	&	4.7			\\														
33	&	27.37	&	29.35	&	4.6			\\												
32	&	27.38	&	29.36	&	4.6			\\														
6	&	27.42	&	29.40	&	4.5				\\															
4	&	27.55	&	29.53	&	4.3		\\														
23	&	27.59	&	29.57	&	4.2			\\

\enddata

\tablenotetext{a}{~Average apparent V magnitude in data from UT 2022 December 19 }
\tablenotetext{b}{~Average absolute magnitude }
\tablenotetext{c}{~Effective diameter, in meters, from Equation \ref{diameter}}
\tablenotetext{d}{~Lightcurve period, hour}
\tablenotetext{e}{~Lightcurve range, magnitudes }
\tablenotetext{f}{~Inferred axis ratio in the sky plane, $a/b = 10^{0.4\Delta V}$ }

\end{deluxetable}

\clearpage


\clearpage
\clearpage

\clearpage

\clearpage
\clearpage
\begin{figure}
\epsscale{0.99}
\plotone{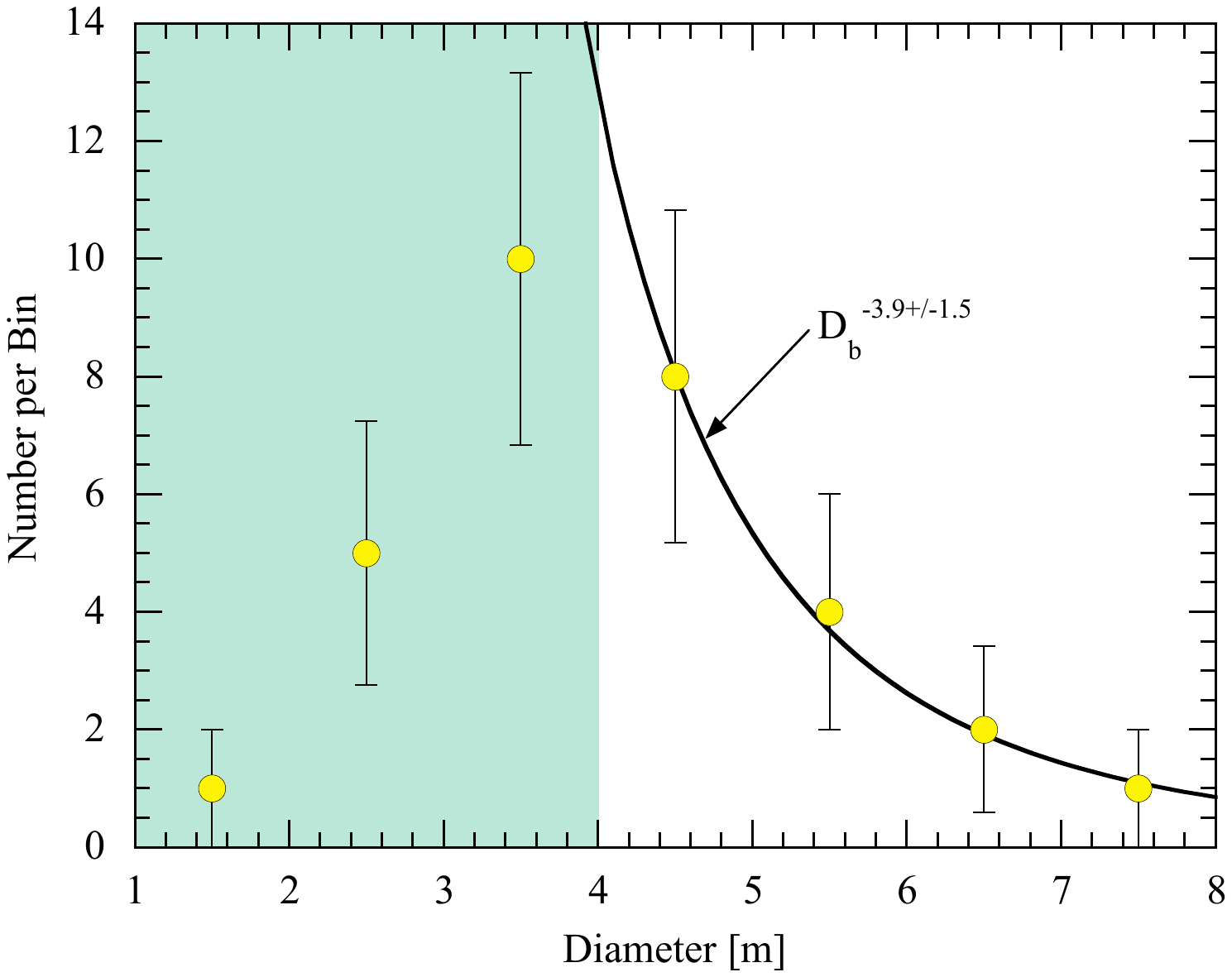}
\caption{The diameter distribution measured from the 36 boulders in Figure \ref{wide} showing roll-over at $D_b \le$ 4 m (shaded background).  A least-squares differential power law fit to the 15 boulders with $D_b >$ 4 m is shown as a solid line, slope $q$ = -3.9$\pm$1.5.
\label{sizes}}
\end{figure}
\clearpage
\begin{figure}
\epsscale{0.73}
\plotone{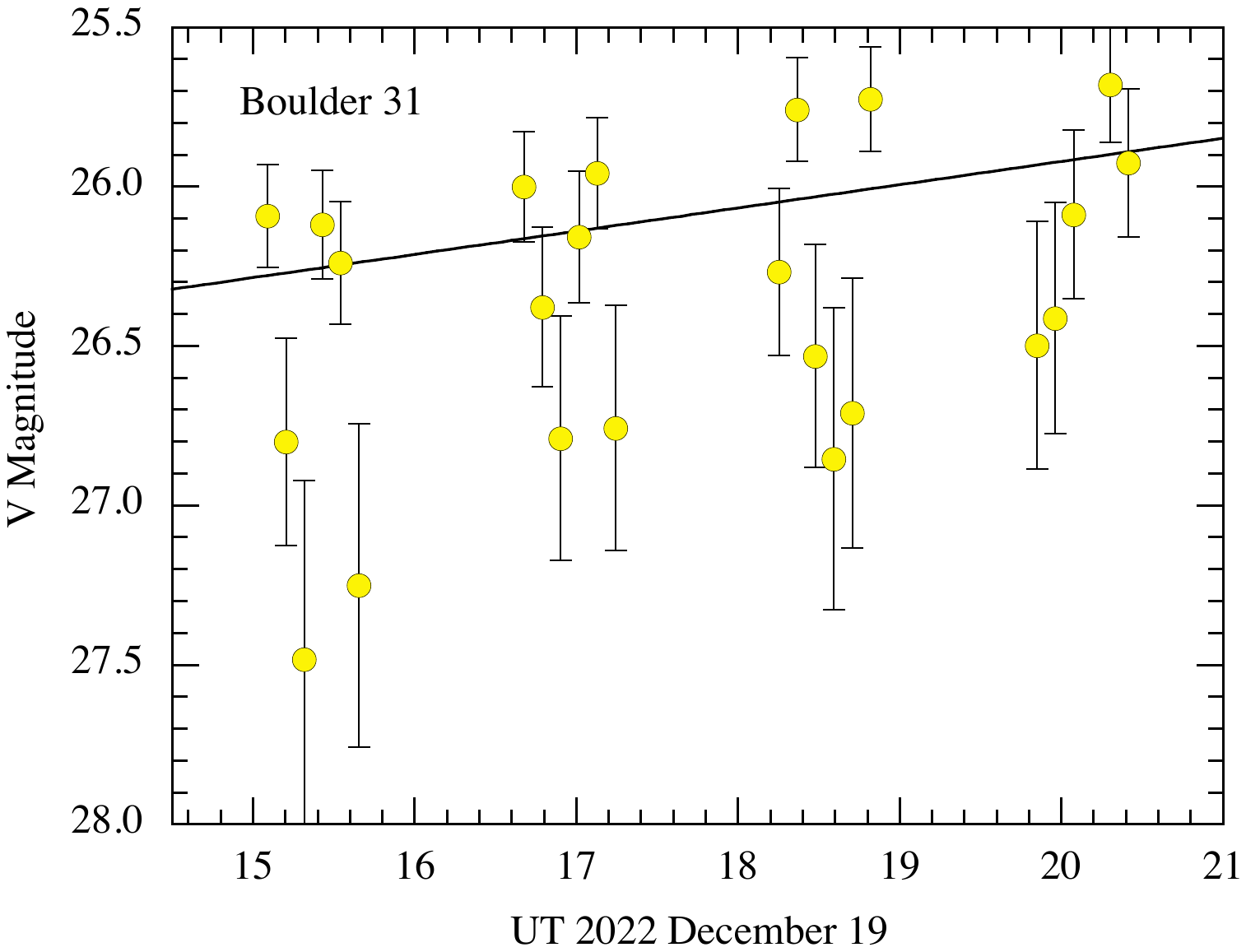}
\plotone{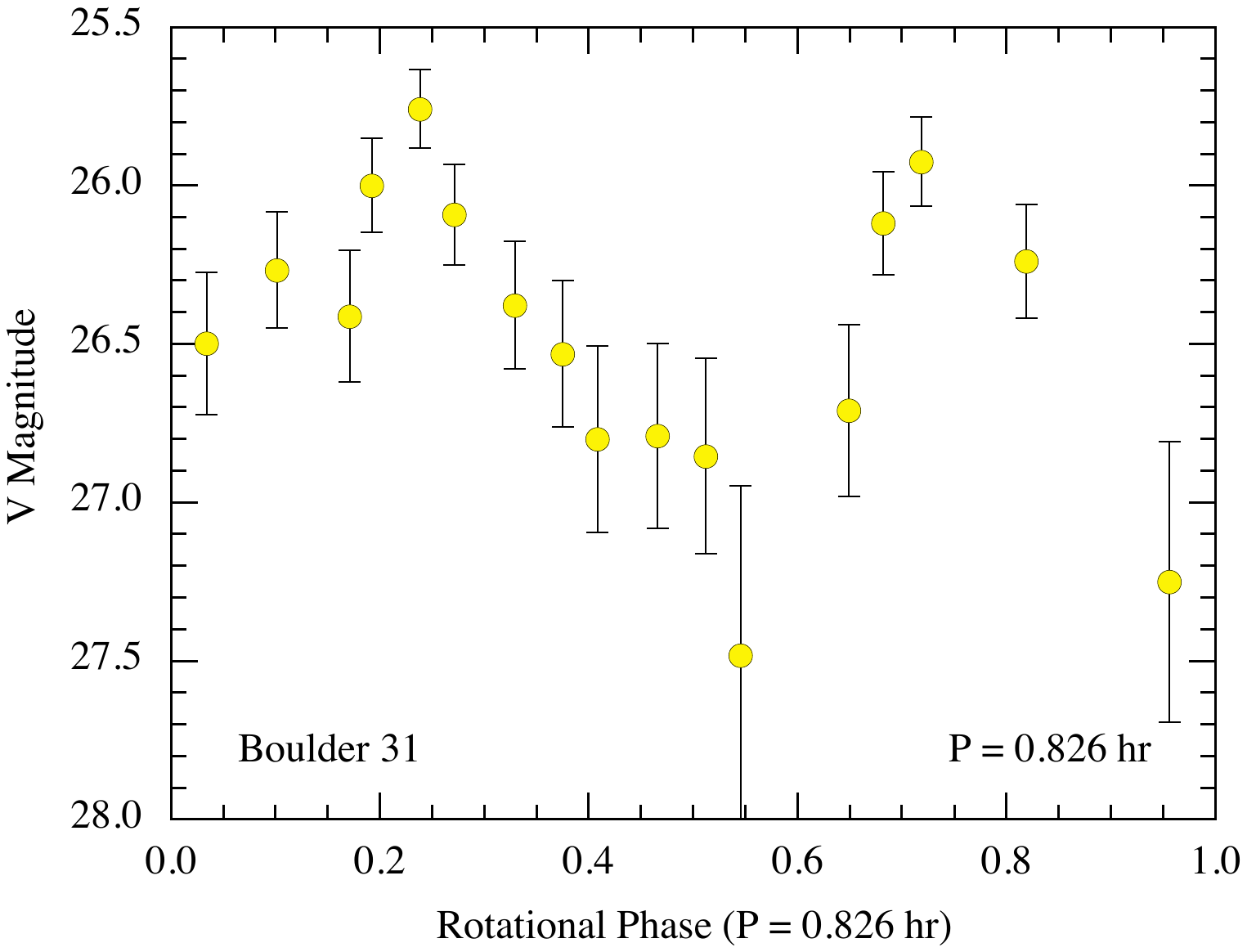}
\caption{(Upper:) Photometry of Boulder 31 showing  large, short-term excursions superimposed on a brightening trend.  The solid black shows a weighted least-squares fit to the data.  (Lower:) Same as upper panel but de-trended and phased to a doubly-periodic lightcurve with period 0.826 hour.  
\label{boulder31}}
\end{figure}
\clearpage
%
%

\begin{figure}
\epsscale{0.99}
\plotone{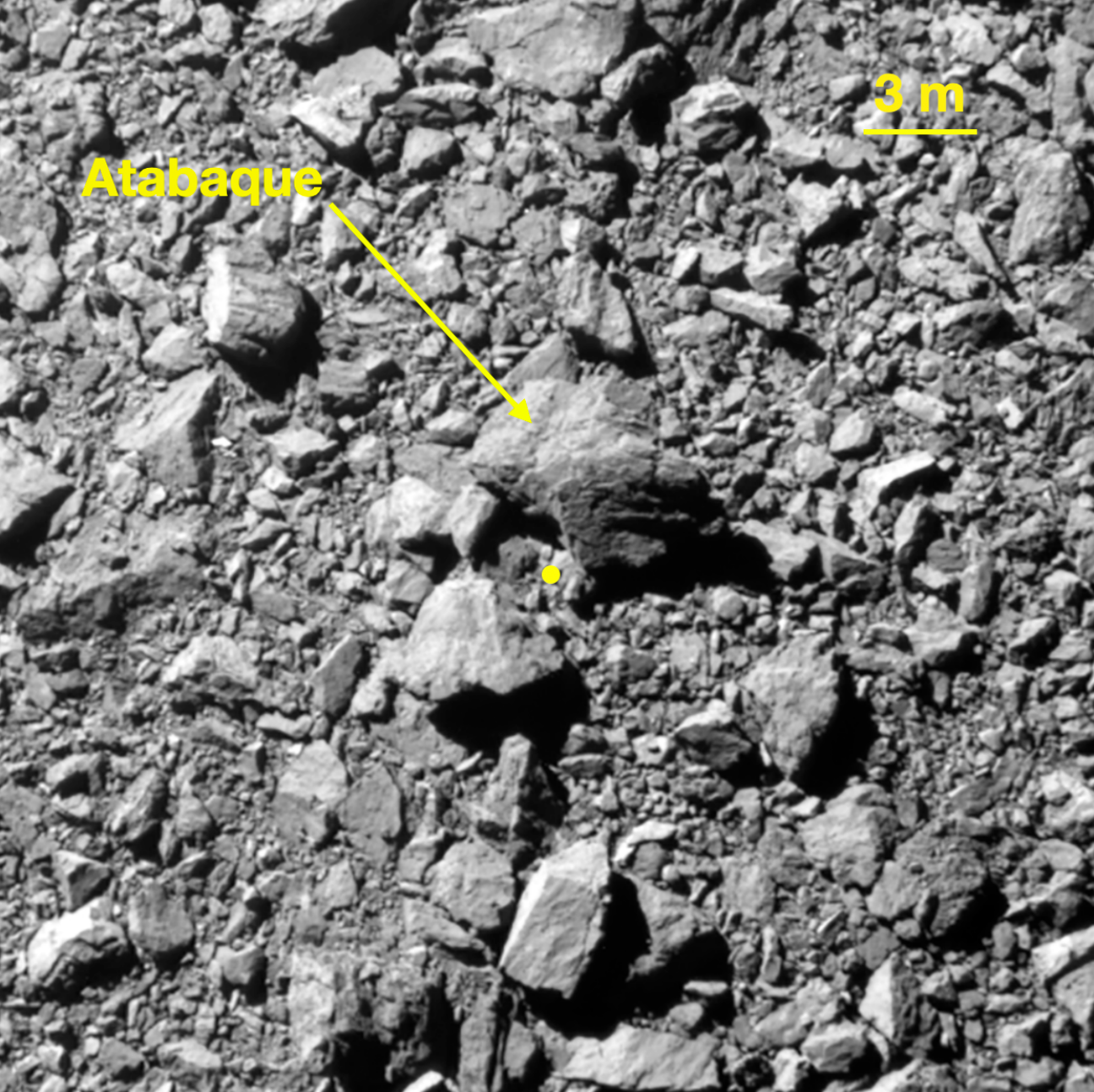}
\caption{The penultimate image recorded from the DART impactor, showing boulders in a field 30 m wide.  The yellow circle marks the nominal DART impact location (Daly et al.~2023) next to the large boulder Atabaque.  Image courtesy NASA/Johns Hopkins APL.\label{penultimate}}
\end{figure}

\clearpage
\begin{figure}
\epsscale{0.99}
\plotone{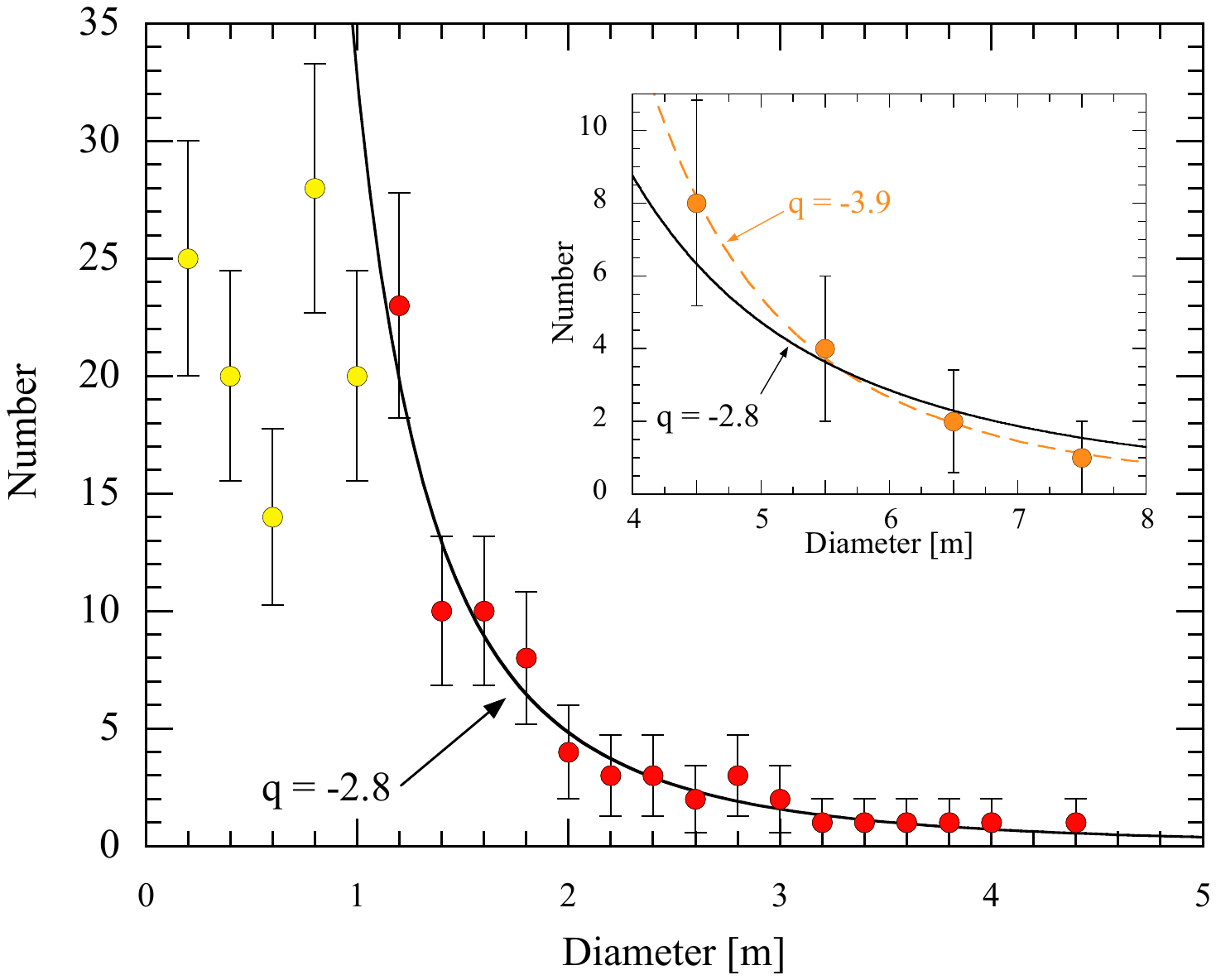}
\caption{Number of boulders per 0.2 m wide bin measured from the image in Figure \ref{penultimate}.  The solid black line shows a weighted least-squares fit to a power law, with differential index $q$ = -2.8$\pm$0.4.  Points shown in yellow (diameters $\le$1 m)  suffer from under-counting and were not used in the fit. The inset shows the diameter distribution inferred from the 15 ejected boulders with $D_b >$ 4 m listed in Table \ref{photometry}.  The dashed orange curve shows the best-fit $q$ = -3.9$\pm$1.5 power law and, for comparison, the solid black line shows $q$ = -2.8 power law deduced from the Didymos surface boulders. \label{boulders}}
\end{figure}

\clearpage
\begin{figure}
\epsscale{0.99}
\plotone{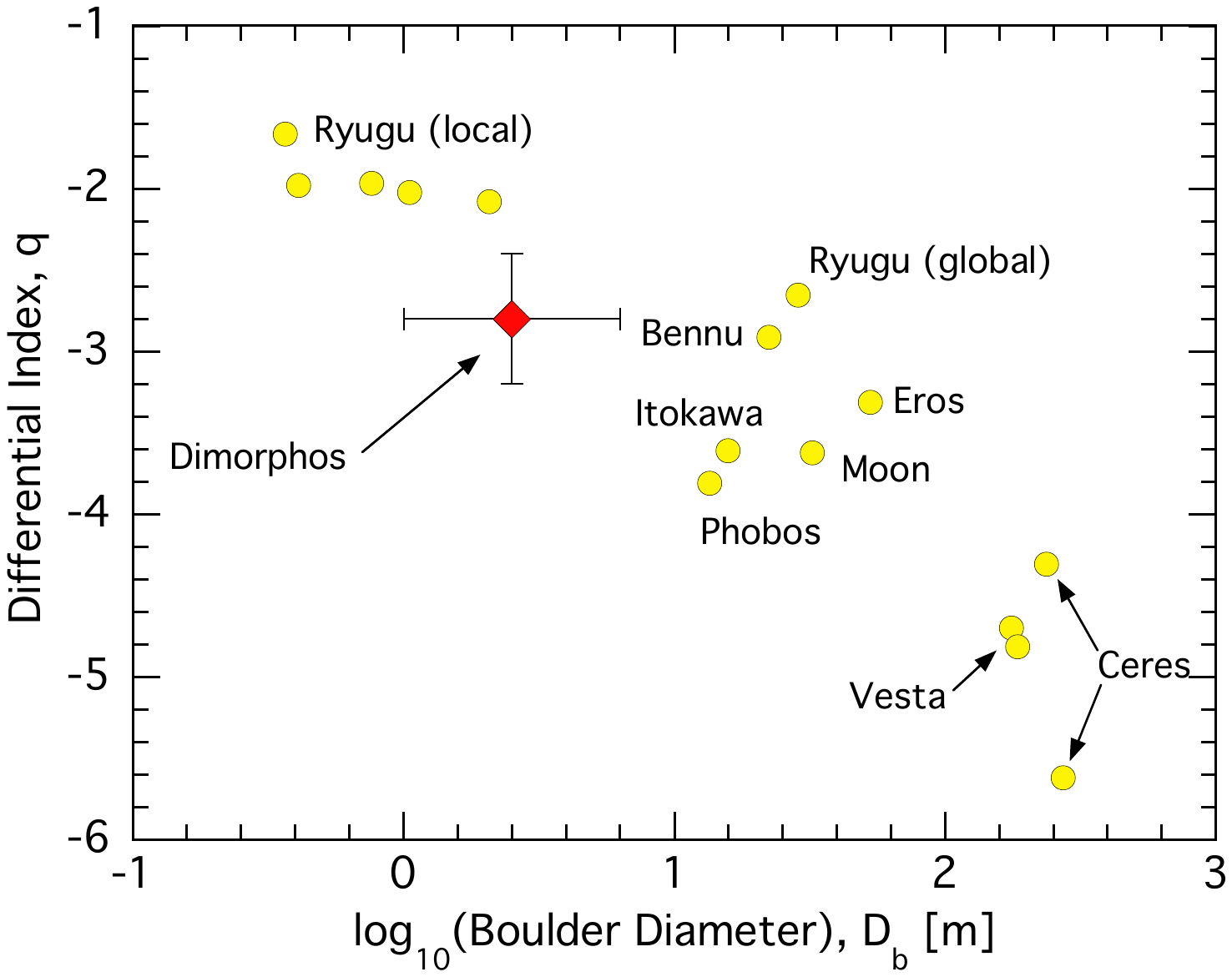}
\caption{Differential power-law indices measured from boulders over a wide range of sizes on different bodies.    The red diamond shows the surface value given in Equation \ref{bouldersize}.  Adapted from Schroder et al.~(2021).
\label{michi}}
\end{figure}

\end{document}